\documentclass[letterpaper,twocolumn,showpacs,prl,aps,floatfix]{revtex4-1}
\usepackage{bm}
\usepackage{multirow,amssymb,amsbsy,amsmath}
\usepackage{graphicx}
\usepackage{braket}

\begin{document}

\newcommand{\tr}{\textrm{Tr}}
\newcommand{\ptr}[1]{\tr_{\textrm{#1}}}
\newcommand{\id}{I}

\title{Detecting nonclassical system-environment correlations by local operations}

\author{Manuel Gessner}

\email{manuel.gessner@physik.uni-freiburg.de}

\author{Heinz-Peter Breuer}

\email{breuer@physik.uni-freiburg.de}

\affiliation{Physikalisches Institut, Universit\"at Freiburg,
Hermann-Herder-Strasse 3, D-79104 Freiburg, Germany}

\date{\today}

\begin{abstract}
We develop a general strategy for the detection of nonclassical system-environment correlations in the initial states of an open quantum system. The method employs a dephasing map which operates locally on the open system and leads to an experimentally accessible witness for genuine quantum correlations, measuring the Hilbert-Schmidt distance between pairs of open system states. We further derive the expectation value of the witness for various random matrix ensembles modeling generic features of complex quantum systems. This expectation value is shown to be proportional to a measure for the quantum discord which reduces to the concurrence for pure initial states.
\end{abstract}


\maketitle

In the theoretical analysis of the dynamics of open quantum systems it is often assumed that the open system $S$ and its environment $E$ are statistically independent at the initial time. Under this assumption the reduced dynamics of the open system can be described in terms of a completely positive dynamical map which operates on the open system's state space, mapping the states of $S$ at time $t=0$ to the states of $S$ at time $t>0$. The one-parameter family of such maps indexed by the time-parameter $t$ then allows the description of Markovian as well as non-Markovian quantum processes \cite{LINDBLAD1976,GORINI,BREUER2007,BLP}.

The mathematical formalism of completely positive transformations provides a fundamental concept in the quantum theory of open systems \cite{BREUER2007} and in quantum information theory \cite{NIELSEN}. However, the assumption of an initially uncorrelated state is in general physically hard to motivate. The description of the reduced system dynamics in terms of linear completely positive maps fails in the case of general initial correlations. Several approaches have been proposed to treat the problem of the dynamics for initially correlated states (see, e.g., \cite{GSI,PECHUKASALICKI,HAYASHI,ROMERO,CARTERET,JORDAN,ROYER,STEL,LINDBLAD1996,RODRIGUEZ,LIDAR}).

One of the central problems in the treatment of correlated systems is given by the fact that the total initial state is typically unknown since a full experimental control of the environment is either impossible or much too complicated in practice, while only the reduced subsystem is experimentally accessible through local measurements and operations. The present work is thus motivated by the question, how can one detect the presence of initial correlations in an unknown system-environment state $\rho$ through only local operations carried out on the open system? Our approach builds on the strategy proposed recently in Ref.~\cite{LPB}, where it is shown that correlations in two total initial states $\rho$ and $\rho'$ can be witnessed by comparison of the time evolution of their reduced open system states $\rho_S(t)$ and $\rho'_S(t)$, respectively. Very recently, experimental realizations of this scheme have been reported \cite{MILAN-EXP,CFL}. In order to witness correlations in a given unknown state $\rho$, the basic idea is to create the second state $\rho'=(\Phi\otimes\id)\rho$ by applying a local trace preserving operation $\Phi$ to the original state $\rho$. Since a local operation can never generate system-environment correlations, any increase of the trace distance between the open system states over its initial value implies the presence of correlations in $\rho$ \cite{LPB}.

The above scheme may fail to detect initial correlations. Whether or not it is successful depends crucially on the chosen local operation and on the structure of the unitary time evolution operator $U_t$ which describes the total system dynamics. Moreover, the general scheme does not explicitly distinguish between classical correlations and genuine quantum correlations of system and environment. The central goals of the present paper are thus twofold. First, we develop a general strategy which allows the detection of true nonclassical correlations, i.e., of correlations described by states with nonzero quantum discord. This is achieved by taking the local quantum operation $\Phi$ to be a dephasing map describing the complete decoherence in a suitable local basis.
The amount to which the evolution of the reduced system states $\rho_S(t)$ and $\rho'_S(t)$ differ from each other can be quantified e.g. by means of the Hilbert-Schmidt distance between these states. Our second goal is to investigate the generic behavior of this distance, without making specific assumptions about the system and its total Hamiltonian. To this end, we determine the expectation value of the Hilbert-Schmidt distance for various random matrix ensembles describing typical features of complex quantum systems.

We consider a state $\rho$ of the total system and its reduced open system state $\rho_S=\ptr{E}\rho$ given by the partial trace over the environment. While the total system state $\rho$ is typically unknown, we assume that the reduced state $\rho_S$ can be determined through state tomography yielding a complete set of rank-one projection operators $\pi_{\mu}$ which project onto the eigenvectors of $\rho_S$. Employing these operators we define a local dephasing operation $\Phi$ acting on operators $A$ of the open system by means of
\begin{equation} \label{MAP-PHI}
 \Phi(A) = \sum\limits_{\mu} \pi_{\mu}A\pi_{\mu}.
\end{equation}
Note that the $\pi_{\mu}$ and the quantum operation $\Phi$ depend on the total system state $\rho$, but the full information necessary to construct these quantities is locally accessible through measurements on the open system. The application of the map $\Phi$ corresponds to a non-selective measurement with respect to the basis given by the projectors $\pi_{\mu}$, and can also be interpreted as describing the complete decoherence of $\rho_S$ in this basis. The corresponding change of the total system state is given by
\begin{equation} \label{DEPHASING}
 \rho \longrightarrow \rho' = (\Phi\otimes\id)\rho
 = \sum\limits_{\mu} \Pi_{\mu}\rho\Pi_{\mu},
\end{equation}
where $I$ denotes the identity map and $\Pi_{\mu} = \pi_{\mu}\otimes I$.

By construction, the marginal states of system and environment are invariant under the application of $(\Phi\otimes I)$, i.e., we have $\rho'_S=\ptr{E}\rho'=\rho_S$ and $\rho'_E=\ptr{S}\rho'=\rho_E$. However, in general the total system state does change under the local dephasing operation. The requirement $\rho=\rho'$ for the invariance of the total state under the dephasing operation can be written as
\begin{equation} \label{CLASSICAL}
 \rho = \sum_{\mu} \Pi_{\mu}\rho\Pi_{\mu}.
\end{equation}
This equation represents the condition for the total system state $\rho$ to have zero quantum discord, i.e., to contain only classical correlations \cite{ZUREK,VEDRAL}. The quantum discord has been introduced as an information theoretic measure for the quantumness of correlations in a composite state $\rho$. The original definition involves two different quantum generalizations of the classical mutual information. Here, we use a technically simpler measure for the degree of quantum correlations as suggested in \cite{LUO}:
\begin{equation} \label{DISCORD}
 \delta(\rho) = \left\|\rho-(\Phi\otimes\id)\rho\right\|,
\end{equation}
where $\|A\|=\sqrt{\tr A^{\dagger}A}$ denotes the Hilbert-Schmidt norm of an operator $A$. The quantity (\ref{DISCORD}) provides a measure for the change of the given state $\rho$ induced by the measurement of the local eigenbasis of the reduced open system state. It vanishes if and only if Eq.~(\ref{CLASSICAL}) holds true, which means that there exists a local measurement basis which does not disturb the total system state, motivating the notion of classical correlations. It is useful to note that the expression (\ref{DISCORD}) can also be written in terms of the purities of the states $\rho$ and $\rho'$. Defining the purity as $\mathcal{P}(\rho)=\tr\{\rho^2\}$ we have
\begin{equation} \label{eq.purity}
 \delta^2(\rho) = \mathcal{P}(\rho)-\mathcal{P}((\Phi\otimes\id)\rho),
\end{equation}
which can easily be verified by use of the relation $\tr\{[(\Phi\otimes\id)\rho]^2\}=\tr\{\rho(\Phi\otimes\id)\rho\}$
which in turn follows from (\ref{DEPHASING}). Thus we see that the measure (\ref{DISCORD}) for quantum correlations is the greater the more the purity is decreased through the local dephasing operation.

Consider now the dynamics of the total system given by some unitary operator $U_t$ on the total system's Hilbert space ${\mathcal{H}}$. We denote by $\rho_S(t)=\ptr{E}\{U_t\rho U_t^{\dagger}\}$ and $\rho'_S(t)=\ptr{E}\{U_t\rho' U_t^{\dagger}\}$ the open system states corresponding to the total initial states $\rho$ and $\rho'=(\Phi\otimes\id)\rho$, respectively. Although the open system states are identical at the initial time $t=0$, the time-evolved states $\rho_S(t)$ and $\rho'_S(t)$ can differ at some later time $t>0$ due to different correlations in the total initial states. This means that different correlations in the initial states become dynamically perceivable in the open system states. If $\rho_S(t)$ and $\rho'_S(t)$ evolve in fact differently one can conclude that the initial state $\rho$ must contain nonclassical correlations since otherwise $\rho$ and $\rho'$ are identical. Thus, we find that the condition
\begin{equation} \label{eq.hilbertschmidt}
 \left\|\ptr{E}\left\{U_t(\rho-\rho') U_t^{\dagger}\right\}\right\| > 0
\end{equation}
implies the presence of nonclassical correlations in the total state $\rho$, i.e., $\delta(\rho) > 0$.

To illustrate the application of our scheme we refer to a recent experiment \cite{CFL} in which initial correlations between the polarization degree of freedom of a photon, representing the open system, and its translational (mode) degree of freedom, forming the environment, have been observed. To detect initial quantum correlations in this experiment one first performs a polarization state tomography on the photons leaving the optical fiber which is used in the experiment to create a correlated initial state. This allows to determine the eigenbasis of the photonic polarization state. The action of the local dephasing map $\Phi$ can then be realized by placing appropriate polarization filters right after the optical fiber (before the photons enter the delay setup), in order to carry out a non-selective measurement of the polarization state in this eigenbasis. The quantity (\ref{eq.hilbertschmidt}) can then be measured by state tomography after the delay setup.

The condition (\ref{eq.hilbertschmidt}) yields a general method for the detection of nonclassical correlations by local quantum operations. However, this condition is only sufficient for the existence of nonclassical correlations since it may happen that
a particular $U_t$ does not lead to an increase of the Hilbert-Schmidt distance in the reduced state space, even though $\rho$ is nonclassically correlated. The question is thus, can one derive general statements about the behavior and the size of the distance in (\ref{eq.hilbertschmidt}) for a generic choice of the time evolution operator? To answer this question we regard the time evolution operator as a unitary matrix $U$ which has been drawn from an appropriate random matrix ensemble over the unitary group $\mathcal{U}(d)$, where $d=d_Sd_E$ with $d_S$ and $d_E$ denoting the dimensions of the Hilbert spaces of the open system $S$ and its environment $E$, respectively. The expectation value of some functional $F(U)$ will be denoted by angular brackets,
\begin{equation}
\langle F(U) \rangle \equiv \int d\mu(U) F(U),
\end{equation}
where $d\mu(U)$ represents the probability measure on the unitary group $\mathcal{U}(d)$. As a first step we consider the matrix ensemble given by the uniform Haar measure of $\mathcal{U}(d)$. Our goal is to determine the average Hilbert-Schmidt distance between the open system states with respect to the Haar measure.

\textit{Theorem.} Let $M$ be an arbitrary fixed Hermitian operator on ${\mathcal{H}}$ and define the reduced system operator
$\Delta=\ptr{E}\left\{UMU^{\dagger}\right\}$. Then we have:
\begin{equation} \label{eq.mainequation}
 \left\langle\left\|\Delta\right\|^2\right\rangle
 = \frac{d_S^2d_E-d_E}{d_S^2d_E^2-1} \left\|M\right\|^2
 +\frac{d_Sd_E^2-d_S}{d_S^2d_E^2-1} \left(\tr M\right)^2.
\end{equation}
The proof is given at the end of the paper. To apply the theorem to the present case we set $M=\rho-\rho'$ which is an operator with trace zero. Therefore, the second term of Eq.~(\ref{eq.mainequation}) vanishes and we find:
\begin{equation} \label{eq.localdetection}
 \sqrt{\left\langle\left\|\ptr{E}
 \left\{U(\rho-\rho')U^{\dagger}\right\}\right\|^2\right\rangle}
 = \sqrt{\frac{d_S^2d_E-d_E}{d_S^2d_E^2-1}} \delta(\rho).
\end{equation}
Thus we obtain the remarkable result that the root mean square of the Hilbert-Schmidt distance between the open system states is proportional to the measure (\ref{DISCORD}) for nonclassical correlations in the initial state $\rho$, with the prefactor depending only on the dimensions $d_S$ and $d_E$. While the witness (\ref{eq.hilbertschmidt}) may fail to detect initial quantum correlations for a particular unitary evolution operator, Eq.~(\ref{eq.localdetection}) shows that the expectation value of this witness with respect to randomly drawn unitaries is nonzero if and only if the initial state contains nonclassical correlations.

An important special case of the general result (\ref{eq.localdetection}) is obtained if $\rho=|\Psi\rangle\langle\Psi|$ is a pure state, such that entanglement is the only possible type of correlations. Employing the Schmidt-decomposition $\ket{\Psi}=\sum_i\lambda_i\ket{\varphi_i}\otimes\ket{\chi_i}$
it is easy to see that the local dephasing operation projects onto the local Schmidt basis $\ket{\varphi_i}$, which yields $\rho'=\sum_i\lambda_i^2\ket{\varphi_i}\!\bra{\varphi_i}
\otimes\ket{\chi_i}\!\bra{\chi_i}$. Using again Eqs.~(\ref{eq.localdetection}) and (\ref{eq.purity}) we find
\begin{equation} \label{pure-state}
 \sqrt{\left\langle\left\|\ptr{E}
 \left\{U(\rho-\rho')U^{\dagger}\right\}\right\|^2\right\rangle}
 = \sqrt{\frac{d_S^2d_E-d_E}{2(d_S^2d_E^2-1)}} C(\rho),
\end{equation}
where $C(\rho)$ denotes the generalized concurrence of $\rho$ \cite{WOOTERS}. This shows that the average distance between the reduced system states is proportional to the concurrence, i.e., to the amount of entanglement in the initial state. We note that Eqs.~(\ref{eq.localdetection}) and (\ref{pure-state}) are not intended to be used in actual experiments to determine $\delta(\rho)$, because the left-hand sides of these equations involve an average over all unitaries which is hard to realize experimentally.

The result (\ref{eq.localdetection}) for the average increase of the Hilbert-Schmidt distance between open system states can be generalized to other physically relevant random matrix ensembles describing generic features of complex quantum systems. To this end, we write the unitary time evolution operator $U_t=\exp\{-iHt\}$ of the total system in terms of a unitary matrix $W$, formed by the eigenvectors of the Hamiltonian $H$, and a diagonal matrix $D$, containing the eigenvalues $E_j$ of $H$, as $U_t=W\exp\{-iDt\}W^{\dagger}$. We assume that the eigenvectors of $H$ are random such that $W$ becomes a random unitary matrix distributed according to the Haar measure.
The characteristic behavior of a generic complex quantum system is determined by the distribution of its level spacing (see, e.g., \cite{HAAKEMEHTA}). In the following we determine the average over $W$, working with a general level spacing distribution which includes, in particular, the cases of regular and chaotic level statistics.

Our aim is to determine the quantity on the left-hand side of Eq.~(\ref{eq.localdetection}) with $U$ replaced by $W\exp\{-iDt\}W^{\dagger}$. The averaging with respect to the Haar measure now requires the determination of an 8th moment of $W$.
The calculations can be performed efficiently by employing the techniques of Ref.~\cite{COLLINS} which are based on advanced
group theoretic methods and the Schur-Weyl duality. The result may be written in the form
\begin{equation}
 \sqrt{\left\langle\left\|\ptr{E}
 \left\{We^{-iDt}W^{\dagger}
 (\rho-\rho')We^{iDt}W^{\dagger}\right\}
 \right\|^2\right\rangle} = c \delta(\rho),
\end{equation}
where the prefactor $c$ is independent of $\rho$, and depends only on the dimensions $d_S$ and $d_E$, and on time $t$ via the Fourier transform $f(t)=\frac{1}{d}\sum_j e^{-iE_jt}$ of the level density. We conclude that, independently of the level spacing distribution, the average Hilbert-Schmidt distance between the open system states is proportional to the measure for quantum correlations given by Eq.~(\ref{DISCORD}). Hence, also for more realistic random matrix ensembles the reduced system distance increases on average if and only if there are initial quantum correlations. The features of specific matrix ensembles, describing e.g. regular or chaotic level statistics, can be investigated in further detail by determining the corresponding averages of the function $f(t)$. Recently, a similar approach has been pursued for the investigation of non-Markovian complex open systems \cite{PINEDA}.

In summary, we have developed a method to locally detect initial correlations in an open quantum system. The witness is successful on average if and only if the correlations are of nonclassical nature. This result nicely complements the recent results of Ref.~\cite{LIDAR} which show that a completely positive dynamical map describing the evolution of the reduced system can be found if and only if the total initial state has zero quantum discord. The method developed here is based on non-selective measurements and state tomography of the reduced system at different times, and should thus be realizable with present-day experimental technologies \cite{MILAN-EXP,CFL}. Finally, we mention that the results of this work are not restricted to an open system setup, the method is applicable to any bipartite system. This includes typical communication protocols, where quantum states are shared by two parties. The correlations of the state can be detected locally by either one of the parties with the aid of the proposed procedure.

\textit{Proof of the theorem.} Let $M=\sum_i m_i\ket{i}\!\bra{i}$ be the spectral decomposition of $M$, and $\left\{\ket{\varphi_i}\right\}$ an orthonormal basis of the open system's Hilbert space $\mathcal{H}_S$. The matrix elements $\Delta_{kl} = \braket{\varphi_k|\Delta|\varphi_l}$ can be expressed as
\begin{equation}
 \Delta_{kl} = \sum_im_i\bra{i}U^{\dagger}\left(\ket{\varphi_l}\!\bra{\varphi_k}\otimes\id\right)U\ket{i},
\end{equation}
which yields for the average Hilbert-Schmidt distance
\begin{equation}
 \left\langle\left\|\Delta\right\|^2\right\rangle
 =\sum_{k,l}\sum_{i,j}m_im_j\bra{i}\Lambda_{A_{kl}A_{kl}^{\dagger}}\,\big(\ket{i}\!\bra{j}\big)\,\ket{j},\label{eq.intermsofmap}
\end{equation}
where we have introduced the map
\begin{equation}
 \Lambda_{AB}(X)=\int d\mu(U) \, U^{\dagger}AUXU^{\dagger}BU,\label{eq.lambda}
\end{equation}
and defined the operators $A_{kl}=\ket{\varphi_l}\!\bra{\varphi_k}\otimes\id$.

\textit{Lemma.} We have
\begin{equation}
\Lambda_{AB}(X)=a\,(\tr X)\id+bX,
\label{eq.maplambda}
\end{equation}
where
\begin{equation}
 a=\frac{d\,\tr BA-\tr A\,\tr B}{d(d^2-1)},\;
 b=\frac{d\,\tr A\,\tr B-\tr BA}{d(d^2-1)}.
\label{eq.aandb}
\end{equation}
Although the proof of this lemma can be carried out with the help of the general method of Ref.~\cite{COLLINS}, here we present an elementary proof. To this end, we use the unitary invariance of the Haar measure to show that the map $\Lambda_{AB}$ has the invariance property $\Lambda_{AB}(WXW^{\dagger})=W\Lambda_{AB}(X)W^{\dagger}$ for all unitary operators $W$. Under the Choi-Jamio{\l}kowski isomorphism \cite{JAMIOL,CHOI} $\Lambda_{AB}$ is mapped to the operator
$\rho_{\Lambda_{AB}}=
\left(\Lambda_{AB}\otimes\id\right)\ket{\Omega}\!\bra{\Omega}$, where $\ket{\Omega}=\frac{1}{\sqrt{d}}\sum_{\omega=1}^d
\ket{\omega}\otimes\ket{\omega}\in\mathcal{H\otimes H}$ is a fixed maximally entangled state. The above unitary invariance of $\Lambda_{AB}$ implies that $\rho_{\Lambda_{AB}}$ is $U\otimes U^*$-invariant or isotropic \cite{VOLLBRECHT,HORODECKI}, with the complex conjugation defined with respect to the basis $\{\ket{\omega}\}$. The general form of such operators is known to be a linear combination of the unit operator and the projection onto the state $\ket{\Omega}$: $\rho_{\Lambda_{AB}}=a\id/d+b\ket{\Omega}\!\bra{\Omega}$. The constants $a$ and $b$ can be obtained by determining the trace of $\rho_{\Lambda_{AB}}$ and the matrix element $\braket{\Omega|\rho_{\Lambda_{AB}}|\Omega}$. Applying the inverse Choi-Jamio{\l}kowski isomorphism to $\rho_{\Lambda_{AB}}$, one finds the general form (\ref{eq.maplambda}) for the map $\Lambda_{AB}$, with $a$ and $b$ given by Eq.~(\ref{eq.aandb}), which proves the lemma.

We now insert the expression (\ref{eq.maplambda}) into Eq.~(\ref{eq.intermsofmap}). Using the relations
$\tr\{A_{kl}^{\dagger}A_{kl}\}=d_E$ and $\tr A_{kl}=\tr A_{kl}^{\dagger}=\delta_{kl}d_E$ one can see that the resulting expression reduces to Eq.~(\ref{eq.mainequation}), which concludes the proof of the theorem.

\end{document}